\documentclass[aps,twocolumn,showpacs,amsfonts,eqsecnum,float,prl]{revtex4}
\usepackage{epsfig,amsmath}
\usepackage{bm}

\begin{document}

\title{Bohmian Mechanics with Complex Action:\\ A New Trajectory-Based Formulation of Quantum Mechanics}

\author{ Yair Goldfarb, Ilan Degani and David J. Tannor} \affiliation{Dept. of
Chemical Physics, The Weizmann Institute of Science,Rehovot, 76100
Israel}

\begin{abstract}
In recent years there has been a resurgence of interest in Bohmian
mechanics as a numerical tool because of its local dynamics, which
suggest the possibility of significant computational advantages for
the simulation of large quantum systems. However, closer inspection
of the Bohmian formulation reveals that the nonlocality of quantum
mechanics has not disappeared --- it has simply been swept under the
rug into the quantum force. In this paper we present a new
formulation of Bohmian mechanics in which the quantum action, $S$,
is taken to be complex. This leads to a single equation for complex
$S$, and ultimately complex $x$ and $p$ but there is a reward for
this complexification --- a significantly higher degree of
localization. The quantum force in the new approach vanishes for
Gaussian wavepacket dynamics, and its effect on barrier tunneling
processes is orders of magnitude lower than that of the classical
force. We demonstrate tunneling probabilities that are in virtually
perfect agreement with the exact quantum mechanics down to $10^{-7}$
calculated from strictly localized quantum trajectories that do not
communicate with their neighbors. The new formulation may have
significant implications for fundamental quantum mechanics, ranging
from the interpretation of non-locality to measures of quantum
complexity.
\end{abstract}
\maketitle


Ever since the advent of Quantum Mechanics, there has been a quest
for a trajectory-based formulation of quantum theory that is exact.
In the 1950's, David Bohm, building on earlier work by
Madelung\cite{madelung} and de Broglie\cite{deBroglie}, developed an
exact formulation of quantum mechanics in which trajectories evolve
in the presence of the usual Newtonian force plus an additional
quantum force\cite{bohm}. In recent years there has been a
resurgence of interest in Bohmian mechanics (BM) as a numerical tool
because of its local dynamics, which suggests the possibility of
significant computational advantages for the simulation of large
quantum systems\cite{courtney,
corey,trahan,jian,erik,sophya,burgha,ginden}. However, closer
inspection of the Bohmian formulation reveals that the non-locality
of quantum mechanics has not disappeared --- it has simply been
swept under the rug into the quantum force. Particularly disturbing
is the fact that for simple cases such as Gaussian wave packet
dynamics of the free particle or the harmonic oscillator, where
classical-quantum correspondence should be perfect, the quantum
force is not only non-vanishing but is the same magnitude as the
classical force\cite{david}.

In this paper we present a new formulation of BM in which the
quantum phase,$S$, is taken to be complex. This leads to a
\textit{single} equation for the complex phase, as opposed to
coupled equations for real phase and real amplitude in the
conventional BM. Complex phase leads to equations of motion for
trajectories with complex $x$ and $p$ but there is a reward for this
complexification --- a significantly higher degree of localization
than in conventional BM. We demonstrate tunneling probabilities that
are in virtually perfect agreement with the exact quantum mechanics
down to $10^{-7}$ calculated from strictly localized quantum
trajectories that do not communicate with their neighbors.  There is
a superficial similarity with some earlier work\cite{boiron,huber2}
on a time-dependent extension of WKB, but the present approach,
which we call Bohmian mechanics with complex action (BOMCA) is
formally exact, and not semiclassical.

The starting point of conventional BM formulation (in 1-dimension)
is the insertion of the ansatz
\begin{equation}
\label{trial}
\psi(x,t)=A(x,t)\exp{\left[\frac{i}{\hbar}S(x,t)\right]},
\end{equation}
in the time dependent Schr\"odinger equation (TDSE), where $A(x,t)$,
$S(x,t)$ are \textit{real} functions representing the amplitude and
phase respectively. Separating the result into its real and
imaginary parts, two PDE's are obtained
\begin{eqnarray}
\label{st0}
S_{t}+\frac{S_{x}^{2}}{2m}+V&=&\frac{\hbar^{2}}{2m}\frac{A_{xx}}{A}, \\
\label{at0}
 A_{t}+\frac{1}{m}A_{x}S_{x}+\frac{1}{2m}AS_{xx}&=&0,
\end{eqnarray}
where $V(x)$ is the potential of the system. The first equation is
referred to as the quantum Hamilton-Jacobi (HJ) equation; it differs
from the classical HJ equation (the LHS) by the addition of a
"quantum potential" $Q\equiv-\frac{\hbar^{2}}{2m}\frac{A_{xx}}{A}$.
Defining a velocity field $v(x,t)=S_{x}(x,t)/m$ the classical HJ
equation yields Newton's equation of motion; the same process for
the quantum HJ equation yields equations of motion for "quantum
trajectories". Eq.(\ref{at0}) can be reformulated as a
hydrodynamic-like continuity equation for probability flow, hence
eqs.(\ref{st0}) and (\ref{at0}) are referred to as the hydrodynamic
formulation of quantum mechanics. The solution of the quantum
hydrodynamic equations along the quantum trajectories constitutes
the conventional BM formulation.

The starting point of the BOMCA formulation is the insertion of the
ansatz\cite{david}
\begin{equation}
\label{trial2}
 \psi(x,t)=\exp{\left[\frac{i}{\hbar}S(x,t)\right]},
\end{equation}
in the TDSE, where we allow the phase to be \textit{complex}. This
yields a \textit{single} newly defined quantum HJ equation
\begin{equation}
\label{st1}
 S_{t}+\frac{1}{2m}S^{2}_{x}+V=\frac{i\hbar}{2m}S_{xx},
\end{equation}
where $Q\equiv-\frac{i\hbar}{2m}S_{xx}$ is the new quantum
potential. Note that there is no expansion in powers of $\hbar$ in
the derivation, hence eq.(\ref{st1}) is an \textit{exact}
formulation of the TDSE which to the best of our knowledge has not
been explored in the literature.

In the spirit of conventional BM our aim is to solve eq.(\ref{st1})
in the Lagrangian approach, that is, along quantum trajectories. A
quantum trajectory is defined by
\begin{equation}
\label{dxdt1}
 \frac{dx}{dt}=v(x,t); \ \ \ v(x,t)\equiv\frac{1}{m}S_{x}(x,t).
\end{equation}
Due to the definition of $x$ as time dependent in eq.(\ref{dxdt1})
we write the solutions of this equation as $x(t;x_{0})$ where
$x_{0}$ is the starting point of the trajectory. Unlike conventional
BM, the complex value of $S$ yields quantum trajectories
$x(t;x_{0})$ that evolve in the complex plane. As a consequence, the
new formulation requires analytic continuation of the wavefunction
and the phase to the complex plane. We consider only analytic
potentials $V(x)$ therefor $v(x,t)$ is analytic in regions which do
not contain nodes of $\psi(x,t)$. To obtain an equation of motion
for $v(x,t)$ we take the spatial derivative of eq.(\ref{st1}) and
apply eq.(\ref{dxdt1}) to obtain
\begin{equation}
\label{mdv1}
 m(v_{t}+vv_{x})-\frac{i\hbar}{2}v_{xx}=-V_{x}(x).
\end{equation}
Identifying the expression in the round brackets as a Lagrangian
time derivative $(\frac{d}{dt}=\frac{\partial}{\partial
t}+v\frac{\partial}{\partial x})$ of $v$, transforms eq.(\ref{mdv1})
to a Newtonian-like equation of motion for the velocity
\begin{equation}
\label{dv}
\frac{dv[x(t;x_{0}),t]}{dt}=\underbrace{-\frac{V_{x}}{m}}_{F_{c}/m}+\underbrace{\frac{i\hbar}{2m}v_{xx}}_{F_{q}/m},
\end{equation}
where we identify $F_{c}$, $F_{q}$ as the classical and the quantum
force respectively. The non-locality of quantum mechanics is
manifested in the appearance of $v_{xx}$ in the quantum force. This
term prevents the first equation in (\ref{dxdt1}) and eq.(\ref{dv})
from being a closed set.

As in conventional BM, the main difficulty lies in estimating the
quantum force. We tackle this problem by taking iterated spatial
partial derivatives of eq.(\ref{dv}). The result can be written
after a short manipulation as
\begin{equation}
\label{set_dvdt}
 \frac{dv^{(n)}}{dt}=-\frac{V^{(n+1)}}{m}+\frac{i\hbar}{2m}v^{(n+2)}-\tilde{g}_{n}
  ; \ \ n=0,...\infty,
 \end{equation}
where $\tilde{g}_{0}=0$ and
$\tilde{g}_{n}=\sum_{j=1}^{n}\binom{n}{j}v^{(j)}v^{(n-j+1)}$ for
$n\geq 1$. The superscripts denote the order of a partial spatial
derivative. The set of eqs.(\ref{set_dvdt}) and the first equation
in (\ref{dxdt1}) are now an infinite but closed set that describes a
\textit{fully local} complex quantum trajectory. If these equations
tend to $0$ as $n\rightarrow\infty$, we may obtain a numerical
approximation by truncating the infinite set at some $n=N$, thus
replacing eq.(\ref{dv}) with a system of $N+1$ coupled ODEs. Since
each individual equation in (\ref{set_dvdt}) depends on the
consecutive $v^{(n+2)}$, the truncation is done by setting
$v^{(N+1)}=v^{(N+2)}=0$. The initial conditions for the $v^{(n)}$'s
are given by
\begin{equation}
\label{initial}
 v^{(n)}(0;x_{0})=\left.\frac{1}{m}\frac{\partial^{n}S_{x}}{\partial
 x^{n}}\right|_{x=x_{0},t=0}=\frac{\partial^{n}}{\partial
 x^{n}}\left.
 \left(-i\hbar\frac{\psi_{x}}{\psi}\right)\right|_{x=x_{0},t=0},
\end{equation}
where we applied the definition from (\ref{dxdt1}) together with
$S=-i\hbar\ln{\psi}$ that follows from eq.(\ref{trial2}). $x_{0}\in
\textbf{C}$ is an initial position of an arbitrary single
trajectory. A similar iterative differentiation process was applied
in conventional BM\cite{jian,trahan,corey} yielding a more
complicated set of coupled equations of amplitude and phase
derivatives.

The relation $v=S_{x}/m$ identifies the \textit{phase field} with an
\textit{action field} of the quantum trajectories. The equation of
motion for the action along a trajectory is similar to its classical
counterpart with the addition of the quantum potential
\begin{eqnarray}
\label{dsdt1}
\frac{dS[x(t;x_{0}),t]}{dt}=S_{t}+vS_{x}=\frac{1}{2}mv^{2}-V+\frac{i\hbar}{2}v_{x}.
\end{eqnarray}
Having $x,v $ and $v_{x}$, the action along a trajectory is obtained
simply by adding the integral of eq.(\ref{dsdt1}) to the initial
value $S(x_{0},0)=-i\hbar \ln{[\psi(x_{0},0)]}$. Inserting the
action in eq.(\ref{trial2}) yields the value of the wavefunction
$\psi[x(t;x_{0}),t]=\exp{\{\frac{i}{\hbar} S[x(t;x_{0}),t]}\}$ at
position $x(t;x_{0})$ in the complex plane.

To obtain the wavefunction on the real axis at time $t_{f}$, in
principle we need to propagate a set of initial positions
$\{x_{0_{j}}\}$ such that $\{x_{j}(t_{f};x_{0_{j}})\}\in \textbf{R}$
at a specified time $t_{f}$. Tracing back the initial positions from
final positions resembles the computationally expensive "root
search" problem familiar from the semiclassical literature. However,
here a set of initial positions can be readily obtained. Suppose the
initial positions are restricted to a region where
$\frac{\partial}{\partial x_{0}}x(t_{f};x_{0})\neq0$ and that the
mapping of initial positions $x_{0}$ to final positions
$x(t_{f};x_{0})$ is an analytic function. Then the inverse mapping
$x\mapsto x_{0}(t_{f};x)$ is also analytic.  Consequently we can
write
\begin{equation}
\label{map}
 x_{0}(t_{f},x_{b})=x_{0}(t_{f};x_{a})+\int_{x_{a}}^{x_{b}}\frac{\partial x_{0}(t_{f};x')}{\partial
 x'}dx',
\end{equation}
where we are free to choose the final positions $x_{a},x_{b}$ and
the integration contour. For simplicity, suppose we have found an
initial condition $x_{0}$ such that $x_{a}=x(t_{f};x_{0})\in
\textbf{R}$. Varying $x_{b}\in \textbf{R}$ and choosing the
integration contour to be the real interval between $x_{a}$ and
$x_{b}$, we obtain from eq.(\ref{map}) a curve $x_{0}(t_{f};x_{b})$
of initial conditions that reach the real axis at $t_{f}$. This can
be translated to a numerical scheme for generating initial positions
that map to the vicinity of the real axis, by writing an iterative
discrete equation based on eq.(\ref{map})
\begin{equation}
\label{mapd}
 x_{0_{j+1}}=x_{0_{j}}+\frac{\delta x_{0_{j}}}{\delta x_{j}}\Delta
 x'; \ \ \ j\geq 1
\end{equation}
where we define $\delta x_{0_{j}}\equiv x_{0_{j}}-x_{0_{j-1}}$ and
$\delta x_{j}\equiv
x_{j}(t_{f};x_{0_{j}})-x_{j-1}(t_{f};x_{0_{j-1}})$. $\Delta x'$ is a
small step along the real axis. Since the final positions are not
exactly on the real axis an \textit{interpolation} process is used
to extract the complex phase along the real axis from the action
values at the set of final positions $\{x_{j}(t_{f};x_{0_{j}})\}$.

\begin{figure}[h]
\begin{center}
\epsfxsize=8.1 cm \epsfbox{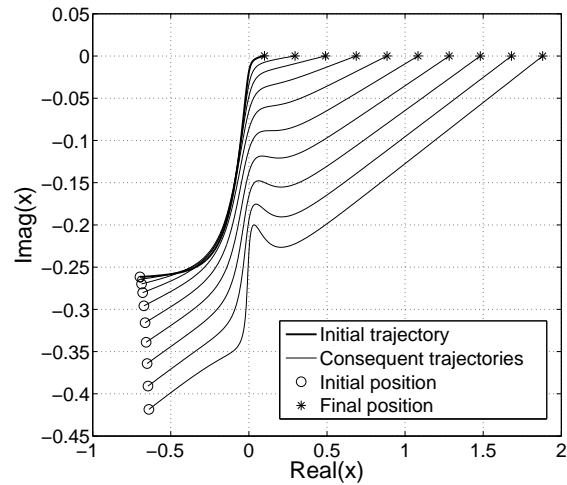} 
\end{center}
\caption{\label{tra} Several complex quantum trajectories for the
scattering of a Gaussian from an Eckart barrier centered around
$x=0$ where $E=p_{c}^{2}/m=0$, $t=1$ and $N=1$. The trajectories
were obtained through the numerical scheme described in the text
(eq.(\ref{mapd})). Note that the trajectories initiate at
$Re(x_{0})\simeq-.7=x_{c}$ and reach $Re[x_{f}(x_{0};t_{f})]>0$,
$Im[x_{f}(x_{0};t_{f})]\simeq 0$. Hence, these trajectories "tunnel"
through the barrier.}
\end{figure}

\begin{figure}[h]
\begin{center}
\epsfxsize=8 cm \epsfbox{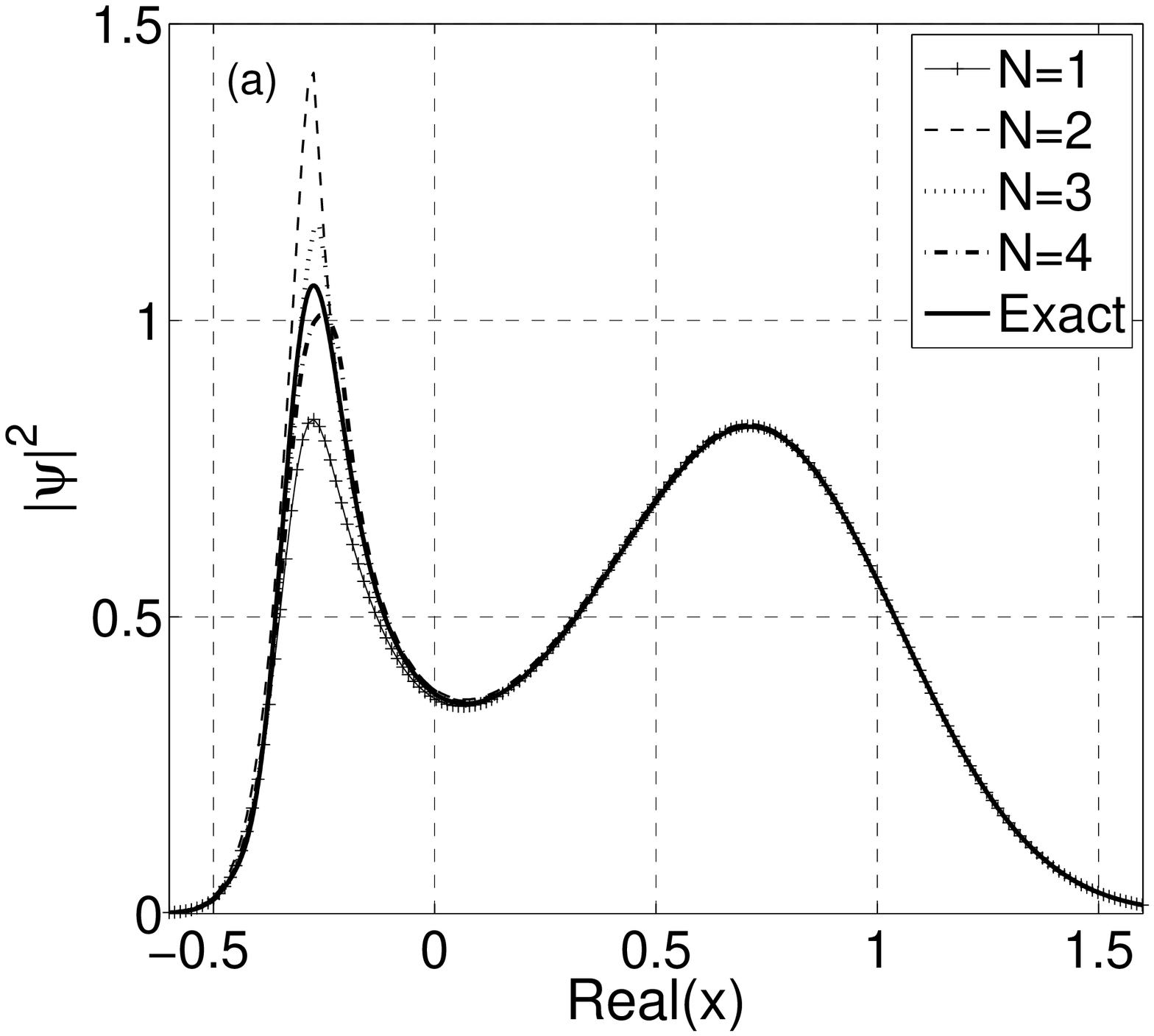} 
\epsfxsize=9 cm \epsfbox{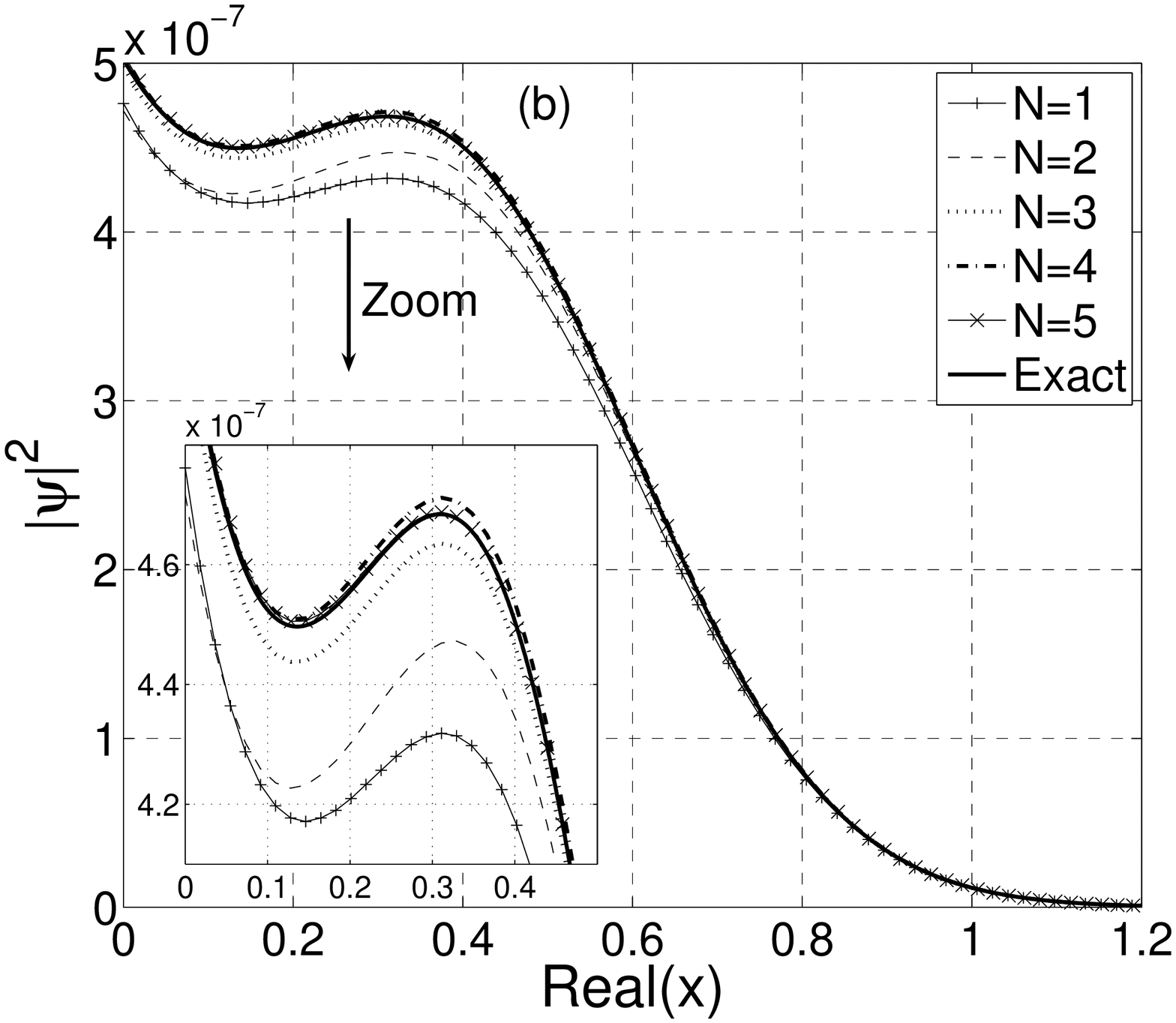} 
\end{center}
\caption{\label{wave3000} Exact wavefunction vs. BOMCA reconstructed
wavefunction for the scattering of a Gaussian from an Eckart
barrier. Plot (a) corresponds to $t=.85$ and $E=50$. Plot (b)
focuses on the transmitted part of the wavefunction for the case of
extremely deep tunneling
--- $t=1$ and $E=0$. Note the convergence to the exact
wavefunction as $N$ increases.  }
\end{figure}

As a numerical example we consider the one-dimensional scattering of
an initial Gaussian wavepacket
$\psi(x,0)=(2\alpha/\pi)^{1/4}e^{\left[-\alpha(x-x_{c})^{2}+\frac{i}{\hbar}p_{c}(x-x_{c})\right]}$
from an Eckart potential $V(x)=D/ \cosh^{2}(\beta x)$. We take
$x_{c}=-.7$, $\alpha=30\pi$, $D=40$, $\beta=4.32$ and $m=30$ (all
units are atomic units). In Fig.\ref{tra} we depict several complex
quantum trajectories for the case of translational energy
$E=p_{c}^{2}/2m=0$, for $N=1$. Note that the complex values of $x$
and $p$ allow the trajectories to "tunnel" through the barrier
centered at $x=0$. In Fig.\ref{wave3000} (a) we compare the exact
wavefunction at $t=.85$, $E=50$ with the BOMCA results for
truncation at $N=1,...,4$. Note that the transmitted part of the
wavefunction is nearly converged for $N=1$, suggesting that BOMCA
will be very efficient for calculating tunneling probabilities. In
Fig.\ref{wave3000} (b) we consider the case of extremely deep
tunneling --- $E=0$ --- and focus on the transmitted part of the
wavefunction. The method converges uniformly, and even for
truncation at $N=1$ the agreement with the exact results is
excellent. It is interesting to contrast the equations for $N=1$
with classical mechanics. For $N=1$ there is no quantum force: the
equations of motion are precisely the \textit{classical} equations
of motion
\begin{equation}
\frac{dx}{dt}=v; \ \ \frac{dv}{dt}=-\frac{V_{x}}{m},
\label{eq:classical_ham}
\end{equation}
albeit for complex $x$ and $v$.  There is however a nonzero quantum
potential that gives an additional term to the action integral (cf.
eq.(\ref{dsdt1})), where the term $v_{x}$ fulfils
\begin{equation}
\frac{dv_{x}}{dt}=-\frac{V_{xx}}{m}-v_{x}^{2} \label{eq:vx_n=1}
\end{equation}
(eq.(\ref{set_dvdt}) with $n=N=1$). It is interesting to note that
the defining equations for $N=1$ (eqs.
\ref{eq:classical_ham}-\ref{eq:vx_n=1} and eq.(\ref{dsdt1})) have
appeared previously in the literature, in the context of
semiclassical methods \cite{huber2,eric}; we emphasize that here
they emerge only as a convenient truncation to an otherwise exact
quantum formulation.


\begin{figure}[h]
\begin{center}
\epsfxsize=7.7 cm \epsfbox{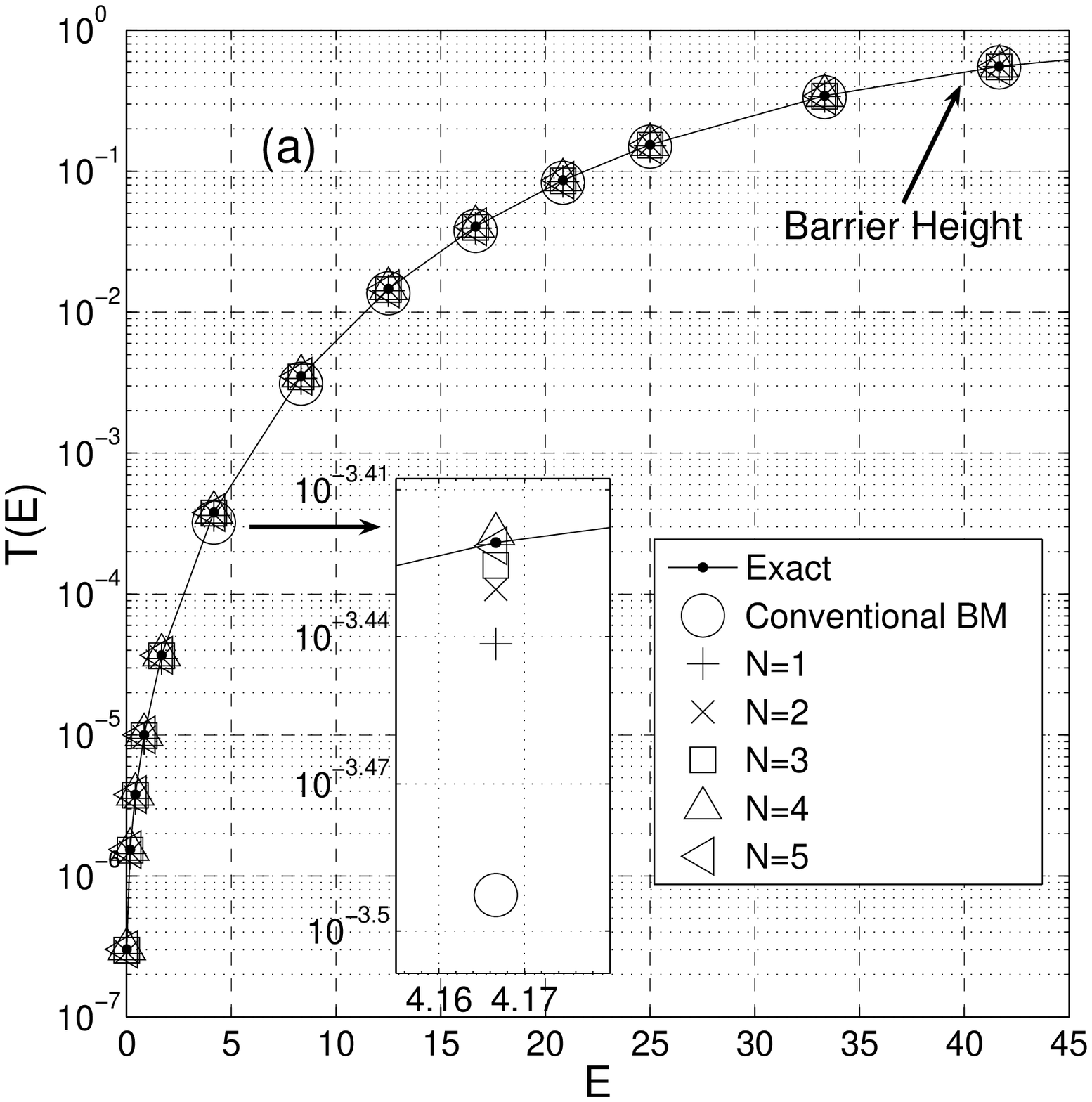} 
\epsfxsize=8.2 cm \epsfbox{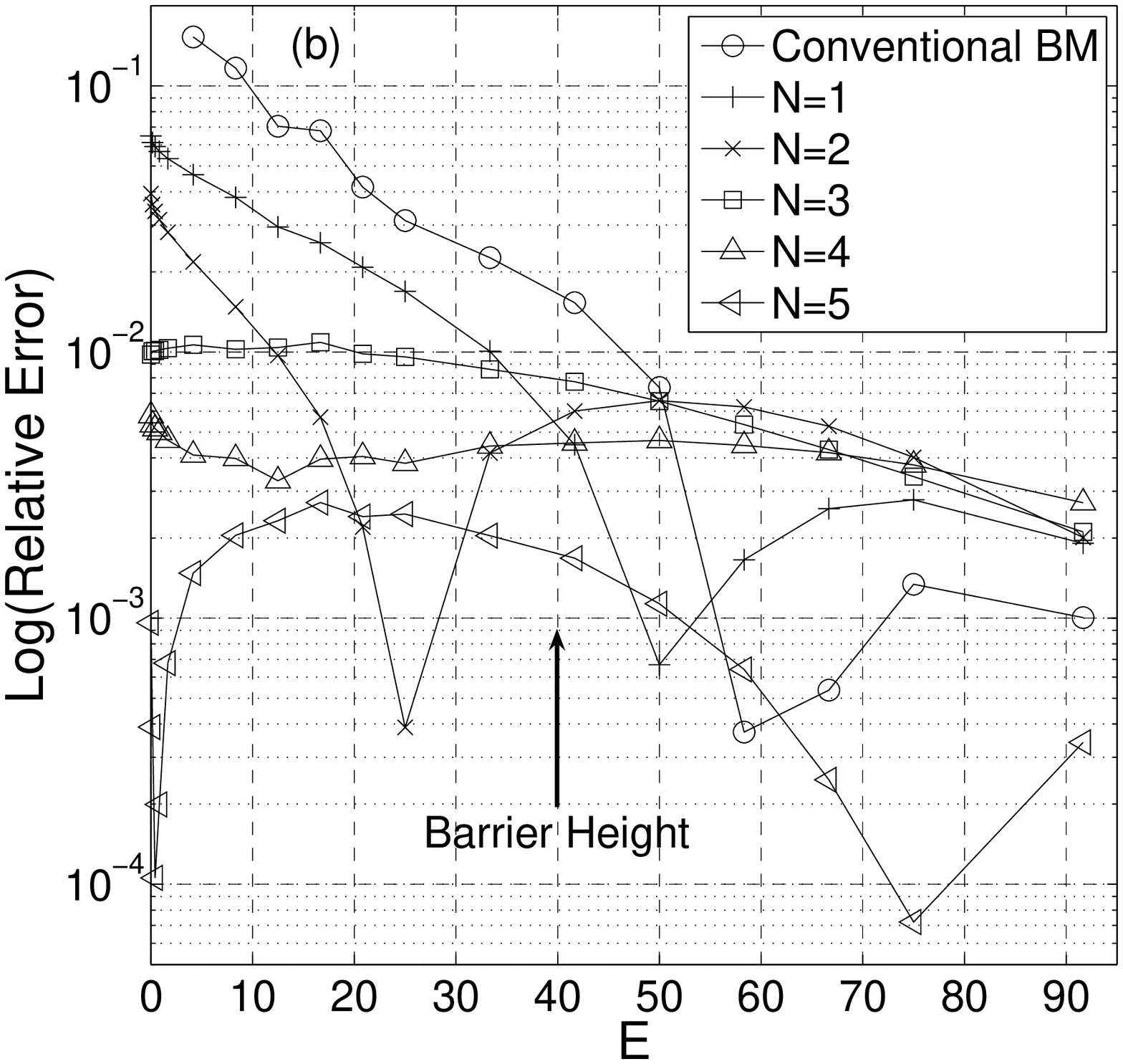} 
\end{center}
\caption{\label{trans} (a) Comparison between the tunneling
probabilities obtained by BOMCA, conventional BM and the exact
results. The inset shows an enlargement of the results for
$E\approx4.17$, the last point for which the BM formulation was
stable; for $E\lesssim4$ we could not obtain stable results from the
conventional BM formulation. (b) Log of the relative divergence from
the exact results. }
\end{figure}

The asymptotic tunneling probability $T(E)$ is calculated by
integrating the absolute square of the wavefunction for $x>0$ at a
sufficiently long time. In Fig.\ref{trans} (a) and (b) we compare
the exact tunneling probabilities as a function of $E$ with the
results obtained from BOMCA and conventional BM. The exact results
were computed by a split operator wavepacket propagation. The BOMCA
results were calculated by propagating 50 complex quantum
trajectories. The conventional BM results were calculated using the
numerical formulation developed by Lopreore and
Wyatt\cite{courtney}. The BOMCA formulation allowed the exploration
of tunneling over the whole energy range, while the conventional BM
formulation proved unstable at low energies ($E\lesssim4$).
Moreover, the BOMCA results are significantly more accurate than the
BM results for all energies below the barrier height $(E<D)$, even
using just the classical equations of motion ($N=1$). Note the
improvement in the accuracy of the BOMCA results as $N$ increases,
suggesting convergence to the exact quantum result. The issue of
convergence will be studied more fully in future work.

In summary we presented BOMCA, a novel formulation of Bohmian
mechanics. This formulation yields simpler equations than
conventional Bohmian mechanics (at the expense of complex
trajectories). Moreover, BOMCA allows a direct and simple derivation
of local uncoupled trajectories that may be used to reconstruct the
wavefunction. The tunneling probabilities obtained by BOMCA for the
scattering process were in excellent agreement with the exact
results even in the extremely deep tunneling regime. We showed that
even classical equations of motion with a small number of (complex)
trajectories are sufficient to obtain very accurate results provided
that an extra, nonclassical term is added to the action integral. We
wish to acknowledge Prof. Edriss S. Titi for several useful
discussions.

\end{document}